# Compatible Learning for Deep Photonic Neural Network

Yong-Liang Xiao[1*], Rongguang Liang[2], Jianxin Zhong[1], Xianyu Su[3], Zhisheng You[3]

[1]*Hunan Key Laboratory for Micro-Nano Energy Materials and Devices, Xiangtan University, Xiangtan, 411105, China*

[2]*College of Optical Sciences, University of Arizona, Tucson, Arizona 85721, USA.*

[3]*National Key Laboratory of Fundamental Science on Synthetic Vision, Sichuan University, Chengdu 610065, China*

*ylxiao@xtu.edu.cn

**Abstract**: Realization of deep learning with coherent optical field has attracted remarkably attentions presently, which benefits on the fact that optical matrix manipulation can be executed at speed of light with inherent parallel computation as well as low latency. Photonic neural network has a significant potential for prediction-oriented tasks. Yet, real-value Backpropagation behaves somewhat intractably for coherent photonic intelligent training. We develop a compatible learning protocol in complex space, of which nonlinear activation could be selected efficiently depending on the unveiled compatible condition. Compatibility indicates that matrix representation in complex space covers its real counterpart, which could enable a single channel mingled training in real and complex space as a unified model. The phase logical XOR gate with Mach-Zehnder interferometers and diffractive neural network with optical modulation mechanism, implementing intelligent weight learned from compatible learning, are presented to prove the availability. Compatible learning opens an envisaged window for deep photonic neural network.

Deep learning has been received as an amazing paradigm for image recognition, language translation, automatic drive and so on [1], it is being implemented in both academia and industry with an explosive growth. The applications strongly require ultrafast speed and little power consumption hardware integrated on a chip that mimicking the biological neurons and synapses, even real-time processing in some occasions. Up to now, there has been some proprietary electronic chips launched for deep learning, such as TrueNorth, TPU, etc. However, the latency in electronic chips generated from transistor computing could induce the limited response speed of information flow and the separate operation in memory and processing [2]. It is well known that, for a long time, photonic neural network can make the light beam individually propagate cross over without crosstalk and has the inherent capability of parallel computation at the speed of light [3-4], which could provide a strong alternative for resolving the interconnection bandwidth limitation and Von Neumann bottleneck in electronic chips. Currently, growing efforts have been directed towards the development of photonic architectures tailored to applications in deep neural network. It is an infancy, but not fantasy.

Photonic neural network has been investigated for several decades within Fourier optics, forward physical architectures of multilayer coherent neural network have anticipated a promising prospect and evolved into on-the-fly Nanophotonics [5]. Recently, deep diffraction neural networks [6] has been applied to validate classification tasks with Terahertz spectrum illumination following Huygens-Fresnel principle, it could approach to massive information capacity with optical analog processing, reaching millions of neutrons and hundreds of billions of connections. Additionally, optical interferometry is also implemented within a reconfigurable nanophotonic chip to ingeniously design an optical intelligent unit [7], forward propagation provides a meshing weight-bank with multi-cascade Mach-Zehner interferometers, the capacity of full connection is somewhat limited to a certain number of ports but it could be broadcasted within nanophotonic circuit. Complex wave in diffraction fulfils the fundamental requirements for superposition in deep neural network and also multi-cascade optical interferometry in nanophotonic chip does. All-optical nonlinearity property of optical material could be explored as a nonlinear synapse in photonic neural network [8], for example, phase change material, $LiNbO_3$, Graphene, III-V etc.

Deep neural network trained successfully on a computer is usually a priori for photonic implementations. Commonly, real-value Backpropagation [9-10] is pervasively used for learning in search for the adjustable parameters. Yet, it is somewhat intractably to understand the nature of phase dynamics and coherent light propagation during training in photonic neural network. In fact, complex-value is always a ubiquitous habitus for coherent wave physical description [11]. The plausible complex-value training with separate format in real part and imaginary part, in which the neurons in the real-part network have dependent connections to those in the imaginary-part network [12], could present vague links with respect to photonic neural network. Thus, efficient learning in complex space for deep photonic neural network is urgently required in a single channel format and easy to be understood and interpreted.

Here we formulate a *compatible learning* protocol in complex space for deep neural network, covering its real counterpart. Particularly, we discover a veiled compatible condition, for the first time, on how to select synapses in complex space, enable learning in real and complex space as a unified model. The selection of nonlinear activation in complex space, has not been unambiguously fabricated to coincide with the learning format. Thus, deep neural network can be successfully trained in real and complex space as a consolidated format, involving weight update rule and nonlinear activation. Assumedly, the compatible learning protocol could be exhaustively suitable and invasively implemented in the deep photonic neural network, and the optical intelligent unit can be designed flexibly with optical matrix manipulation/modulation according to the learned intelligent matrix in complex space.

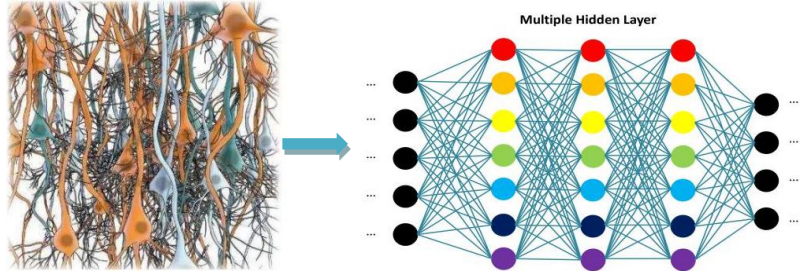

Fig.1 Schematic diagram of deep neural network in complex space mimicked from biological neuron

*Forward propagation* Neural network originates from the neuroscience, the current salient feature of deep learning is that the limited successes for outstanding information processing through hierarchical multilayer networks in complex space [13]. Deep neural network in complex space involves many different linear combiners, each of which has a nonlinearity activation at the outputs. The forward network in complex space can be expressed as pairs of real numbers, but the training presents intractable, and it will be more convenient for analysis and interpretation as a single complex entity. The forward propagation for full connections in complex space is expressed as

$$\mathbf{x}_\ell = f(\mathbf{z}_\ell), \mathbf{z}_\ell = \mathbf{W}_\ell \mathbf{x}_{\ell-1} + \mathbf{b}_\ell \quad (1)$$

*Backward Propagation* Square-error is regarded as the conventional loss function. We implement the gradient operator for finding possible optimize path when process a function of the independent $z$ and $z^*$. The gradient operator can solve the minimization problem directly and mathematically in secure with respect to complex matrices. We can compute the complex variables derivatives of loss energy function with respect to the weight and bias using complex-chain rules. Define $\ell$ denotes the current layer in the multilayer deep neural network in complex space with full connections, and holistic layer is designed to be $L$.

$$\Delta \mathbf{W}_\ell = -\eta \nabla_{\mathbf{w}_\ell^*}\left[\varepsilon\varepsilon^*\right] = -\eta\varepsilon\nabla_{\mathbf{w}_\ell^*}\left[\varepsilon^*\right] = -\eta\varepsilon\nabla_{\mathbf{w}_\ell^*}\left[\mathbf{t}_{L+1}^* - f(\mathbf{W}_{L+1}^* f(\mathbf{W}_L^* \mathbf{x}_{L-1}^* \cdots f\left[\mathbf{W}_{\ell+1}^* f(\mathbf{W}_\ell^* \mathbf{x}_{\ell-1}^* + \boldsymbol{\theta}_\ell^*) + \boldsymbol{\theta}_{\ell+1}^*\right]\cdots ) + \boldsymbol{\theta}_\ell^*) + \boldsymbol{\theta}_{\ell+1}^*)\right]$$

$$\nabla_{\mathbf{w}_\ell^*}\left[\varepsilon^*\right] = \nabla_{\mathbf{w}_\ell^*}\left[\mathbf{t}_{L+1}^* - f(\mathbf{W}_{L+1}^* f(\mathbf{W}_L^* x_{L-1}^* \cdots f\left[\mathbf{W}_{\ell+1}^* f(\mathbf{W}_\ell^* \mathbf{x}_{\ell-1}^* + \boldsymbol{\theta}_\ell^*) + \boldsymbol{\theta}_{\ell+1}^*\right]\cdots ) + \boldsymbol{\theta}_\ell^*) + \boldsymbol{\theta}_{\ell+1}^*)\right]$$

$$= -\nabla_{\mathbf{w}_\ell^*} f\left(\mathbf{net}_{L+1}^*\right)\{\frac{\partial\left[\mathbf{net}_{L+1}^*\right]^T}{\partial \mathbf{x}_L^*}\frac{\partial\left[\mathbf{x}_L^*\right]}{\partial \mathbf{net}_L^*}\}\{\frac{\partial\left[\mathbf{net}_L^*\right]^T}{\partial \mathbf{x}_{L-1}^*}\frac{\partial\left[\mathbf{x}_{L-1}^*\right]}{\partial \mathbf{net}_{L-1}^*}\}\cdots\{\frac{\partial\left[\mathbf{net}_{\ell+2}^*\right]^T}{\partial \mathbf{x}_{\ell+1}^*}\frac{\partial\left[\mathbf{x}_{\ell+1}^*\right]}{\partial \mathbf{net}_{\ell+1}^*}\}\{\frac{\partial\left[\mathbf{net}_{\ell+1}^*\right]^T}{\partial \mathbf{x}_\ell^*}\frac{\partial\left[\mathbf{x}_\ell^*\right]}{\partial \mathbf{net}_\ell^*}\}\frac{\partial\left[\mathbf{net}_\ell^*\right]}{\partial \mathbf{W}_\ell}$$

$$= -f'\left(\mathbf{net}_{L+1}^*\right)\mathbf{W}_L^H f'\left(\mathbf{net}_L^*\right)\mathbf{W}_{L-1}^H f'\left(\mathbf{net}_{L-1}^*\right)\cdots \mathbf{W}_{\ell+2}^H f'\left(\mathbf{net}_{\ell+2}^*\right)\mathbf{W}_{\ell+1}^H f'\left(\mathbf{net}_{\ell+1}^*\right)\mathbf{W}_\ell^H \mathbf{x}_{\ell-1}^*$$

$$\boldsymbol{\delta}_{L+1} = \varepsilon \odot f'\left(\mathbf{net}_{L+1}^*\right), \boldsymbol{\delta}_L = \mathbf{W}_{L+1}^H \boldsymbol{\delta}_{L+1} \odot f'\left(\mathbf{net}_L^*\right), \boldsymbol{\delta}_{L-1} = \mathbf{W}_L^H \boldsymbol{\delta}_L \odot f'\left(\mathbf{net}_{L-1}^*\right)\cdots$$

$$\boldsymbol{\delta}_{\ell+1} = \mathbf{W}_{\ell+1}^H \boldsymbol{\delta}_{\ell+2} \odot f'\left(\mathbf{net}_{\ell+1}^*\right), \boldsymbol{\delta}_\ell = \mathbf{W}_\ell^H \boldsymbol{\delta}_{\ell+1} \odot f'\left(\mathbf{net}_\ell^*\right)$$

$$\Delta \mathbf{W}_\ell = \eta \mathbf{x}_{\ell-1}^* \boldsymbol{\delta}_\ell^T \quad (2)$$

Complex matrix and real matrix representation with Backpropagation, for comparison, are presented as follows

**Complex Space**

$\mathbf{x}_\ell = f(\mathbf{y}_\ell), \mathbf{y}_\ell = \mathbf{W}_\ell \mathbf{x}_{\ell-1} + \mathbf{b}_\ell$

$\begin{cases} \boldsymbol{\delta}_L = f'(\mathbf{y}_L^*) \odot (\mathbf{y}_L - \mathbf{t}_L) \\ \boldsymbol{\delta}_\ell = \left[\mathbf{W}_{\ell+1}^*\right]^T \boldsymbol{\delta}_{\ell+1} \odot f'(\mathbf{y}_\ell^*) \\ \Delta \mathbf{W}_\ell = -\eta \mathbf{x}_{\ell-1}^* \left[\boldsymbol{\delta}_\ell\right]^T \\ \Delta \mathbf{b}_\ell = -\eta \left[\boldsymbol{\delta}_1\right]^T \end{cases}$

$\mathbf{x}_\ell, \mathbf{x}_{\ell-1}, \mathbf{y}_\ell, \mathbf{y}_L, \mathbf{t}_L, \boldsymbol{\delta}_L, \boldsymbol{\delta}_\ell, \Delta \mathbf{W}_\ell, \Delta \mathbf{b}_\ell, \in \mathbb{Z}$

**Real Space**

$\mathbf{x}_\ell = f(\mathbf{u}_\ell), \mathbf{u}_\ell = \mathbf{W}_\ell \mathbf{x}_{\ell-1} + \mathbf{b}_\ell$

$\begin{cases} \boldsymbol{\delta}_L = f'(\mathbf{u}_L) \odot (\mathbf{u}_L - \mathbf{t}_L) \\ \boldsymbol{\delta}_\ell = \left[\mathbf{W}_{\ell+1}\right]^T \boldsymbol{\delta}_{\ell+1} \odot f'(\mathbf{u}_\ell) \\ \Delta \mathbf{W}_\ell = -\eta \mathbf{x}_{\ell-1} \left[\boldsymbol{\delta}_\ell\right]^T \\ \Delta \mathbf{b}_\ell = -\eta \left[\boldsymbol{\delta}_\ell\right]^T \end{cases}$

$\mathbf{x}_\ell, \mathbf{x}_{\ell-1}, \mathbf{u}_\ell, \mathbf{u}_L, \mathbf{t}_L, \boldsymbol{\delta}_L, \boldsymbol{\delta}_\ell, \Delta \mathbf{W}_\ell, \Delta \mathbf{b}_\ell \in \Re$

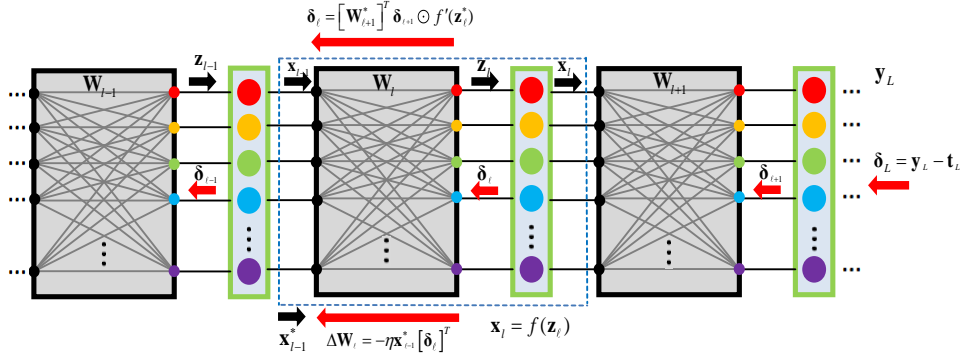

Fig. 2. The flow diagram of compatible learning with backpropagation

The training and activation implementing separate strategy in real and imaginary parts (supplementary materials 1) with real-value Backpropagation is abandoned here [14]. Seen from the formulation comparisons, matrix representation in complex space covers its real counterpart. The update rules involving nonlinear activation for Backpropagation also can degrade into a real counterpart without any ambiguity. It is the *compatibility*. The compatible condition, along with compatible learning during partial derivation with gradient operator, is enforced as $f^*(\mathbf{z}_\ell) = f(\mathbf{z}_\ell^*)$. Under the compatible condition, the prevalent sigmoid and tanh in real space, could be generalized to its complex space by artificially replacing real-variable with complex-variable directly. The essence of compatible learning can be realized within a line of code. Compatible learning is applied very easily and directly for photonic deep learning. The schematic diagram of compatible learning is exhibited in Figure. 2.

The compatible learning is not deduced from C-R(*Cauchy-Riemann)* condition while the C-R condition has been employed to deduce an incomplete matrix representation [15], the complete matrix representation based on the C-R condition for the BackPropogation error update is

$$\boldsymbol{\delta}_\ell = \left[\mathbf{W}_{\ell+1}^*\right]^T \delta_{\ell+1} \odot \left[f'_{C-R}(\mathbf{z}_\ell)\right]^* \quad (3)$$

$f_{C-R}$ is a complex analytic activation function. There is a little discrepancy in Eq(3) from our compatible model with respect to the derivation of nonlinear-activation function. Seen from the Eq(3), both sigmoid and tanh, unfortunately, can't be utilized as nonlinear activations in complex space, because sigmoid and tanh represented in its complex space don't satisfy C-R condition. However, sigmoid and tanh can be quoted directly as nonlinear activations in complex space under the compatible condition, even though these two functions utilized in complex space had been implemented ambiguously for over three decades [16-17]. It is, so, the same situation that separate activation in real and imaginary parts may be designed more elaborately within harmonic analysis of complex function if C-R condition is imposed. Comparing the two matrix representations based on compatible condition and C-R condition, the main difference is with respect to the derivation of nonlinear activation. $f'(\mathbf{z}_\ell^*)$ is a function that real variable is directly replaced by the complex variable after real derivation, while $[f'_{C-R}(\mathbf{z}_\ell)]^*$ is a function that it is derived on the basis of mathematically analytical meaning. Benefiting on the fact that the formats of alternative functions for nonlinear activation are very limited, the underlying relationship between the two matrix representations could be build, associating the intrinsic property $\left[f'_{C-R}(\mathbf{z})\right]^* = \left[f'_{C-R}(\mathbf{z}^*)\right]$ with the compatible condition $f^*(\mathbf{z}) = f(\mathbf{z}^*)$.

$$\begin{cases} f^*(\mathbf{z}) = f(\mathbf{z}^*) \\ \forall f_{C-R}, \left[\dfrac{\partial f_{C-R}(\mathbf{z})}{\partial \mathbf{z}}\right]^* = \dfrac{\partial f_{C-R}(\mathbf{z}^*)}{\partial \mathbf{z}^*} \end{cases}$$

$$\frac{\partial f_{C-R}(\mathbf{z},\mathbf{z}^*)}{\partial \mathbf{z}} = \frac{\partial \left[f_{C-R}(\mathbf{z}^*,\mathbf{z})\right]^*}{\partial \mathbf{z}} = \left[\frac{\partial \left[f_{C-R}(\mathbf{z},\mathbf{z}^*)\right]^*}{\partial \mathbf{z}^*}\right]^*$$

Thus, $\left[f'(\mathbf{z}_\ell^*)\right] = \left[f'_{C-R}(\mathbf{z}_\ell)\right]^*$. In brief, the matrix mode for Backpropagation in complex space has an equivalence if the *compatible condition* and the *C-R condition* are satisfied simultaneously.

The optimization algorithms should be suitably chosen in accordance with different training tasks, such as stochastic gradient descent and its variants [18]. Loss function adapted learning rate $\eta$ etc, will influence the convergence speed and accuracy. Another significant point on which one should focus is the initialization, there are many heuristic proposals with variance invariance during epochs [19-20] so that one can try to explore in the initialization. In the following, we would like to implement compatible learning for deep photonic neural network training.

*Learning for phase logical XOR gate*

In order to test the validity of compatible learning, we numerically demonstrate the training of phase logical XOR gate [21], which is a resemblance with the common logical XOR by phase rotation and radius scaling. The label pairs are cultivated as seed to phase logical XOR gate for training a $4 \to 4$ mapping relationship shown in Table 1. The diagram of phase logical XOR gate is shown in figure 3, two 4×4 intelligent matrix learned from two-layer neural network in complex space are utilized mapping the phase and radius skip. The digital experiments validate compatibility for a perfect consolidated training, even though both real-value and complex-value are mingled together as the training samples. The phase logical XOR gate can be realized by phase rotation, the radius skip can also be trained out accompanying the phase. Gradient descent is the optimization algorithm used during epochs. The complex weight is initialized through uniform statistically distribution with zero mean-value as well as the settled variance in the separate real and imaginary part respectively.

Table 1 The training samples(X │ Y) for phase logical XOR gate (amplitude1 =2, amplitude2 =1)

| Samples | X1 | X2 | X3 | X4 | Y1 | Y2 | Y3 | Y4 |
|---|---|---|---|---|---|---|---|---|
| Phase1 | π/4 | 3π/4 | 5π/4 | 7π/4 | 0 | π/2 | π | 3π/2 |
| Phase2 | π/3 | 5π/6 | 8π/6 | 11π/6 | π/6 | 4π/6 | 7π/6 | 10π/6 |

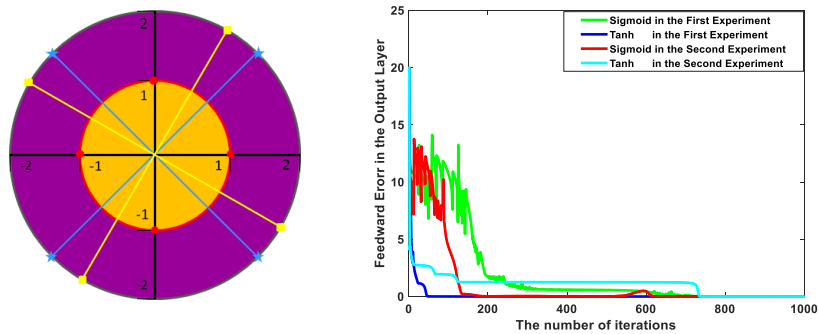

Fig.3 Phase logical XOR gate in complex space and corresponding learning epochs

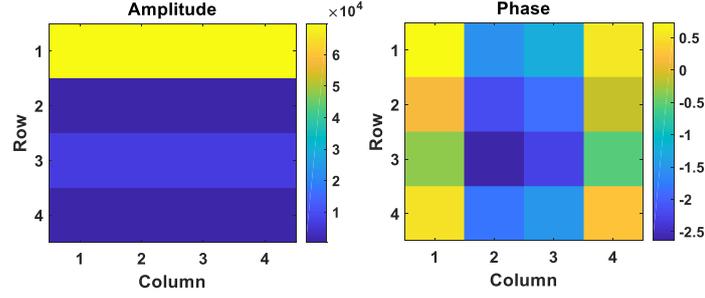

**(a)** The amplitude and phase for the first complex weight matrix

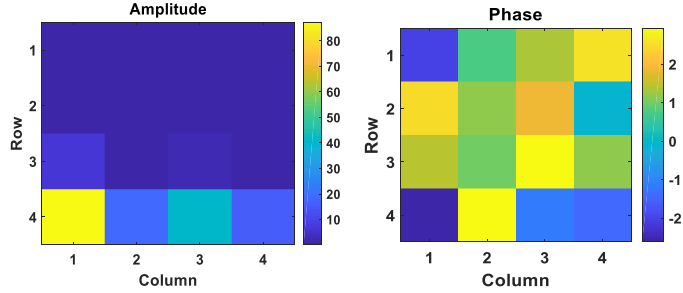

**(b)** The amplitude and phase for the second complex weight matrix

Fig.4 The learned complex weight demonstration for phase logical XOR gate

In the numerical experiments, we exploit two basic nonlinear activation functions sigmoid and tanh to observe the convergent performance. Seen from the curves in figure 3, both sigmoid and tanh can be directly implemented in compatible learning, making the forward output converge to labels very well with two different initialization experiments. In most case of our simulations, the convergence speed of tanh is superior to sigmoid with the same initialization. Our compatible model successfully learns the phase logical XOR gate in around 200 epochs with sigmoid activation and around 20 epochs with tanh activation in the learning curves. The learning rate is constantly settled as 0.1. Figure 4 exhibits the two distributions of the learned complex weight with sigmoid nonlinear activation. Amplitude of the complex weight is close to $10^4$ in the first layer and $10^2$ in the second layer without any regularization, the phase distribution is wrapped between [-π, π] in a period. The detailed value of the learned weight matrix is embedded as the color demonstration. And here the learned weight matrix is not a unitary without Riemannian gradient corrections enforced on the Euclidean complex weight [22]. The phase logical XOR gate indicates that compatible learning is feasible and available. The optical implementation for phase logical XOR gate can be realized in a meshing Mach-Zehner interferometer with four IO ports [23-24], as shown in Figure 5. The optical principle of forward implementation is given in supplementary materials 2.

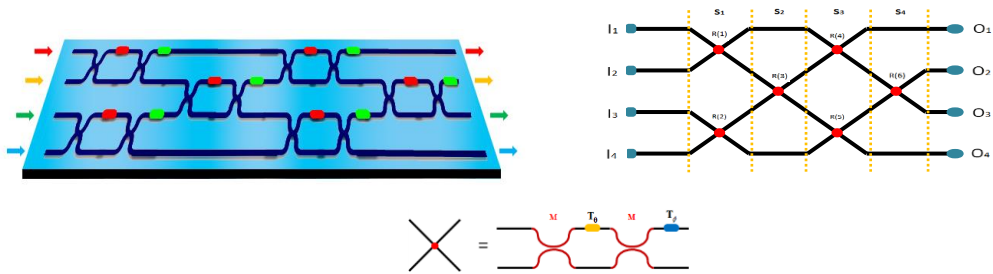

Fig.5 The optical realization of learned complex weight for phase logical XOR gate. 4×4 rectangular MZI optical network structure is reconfigurable by four levels, S1 to S4. Each MZI unit uses 3dB directional coupler, while using two phase shifters to tune phase $\theta$ and $\phi$.

*Learning for diffractive neural network*

Coherent diffraction artificial intelligence has a strong potential for photonic neural network dense interconnections due to its coherent superposition and diffractive distance. Compatible learning protocol is considerably suitable for training diffractive neural network since complex-value represents the natural entity of optical wave in coherent diffraction, even though a single intensity signal is directly recorded in the output layer without holographic interferometry. The learnable weight in complex space at each layer could be considered as an optical intelligent matrix to be modulated. But the optical intelligent matrix is very difficult to directly implement with optical diffraction. We show an optical setup in Figure 6(a), implementing optical matrix-vector multiplication alternative for coherent optical neural network, rendering the optical mask modulation and activation simultaneously.

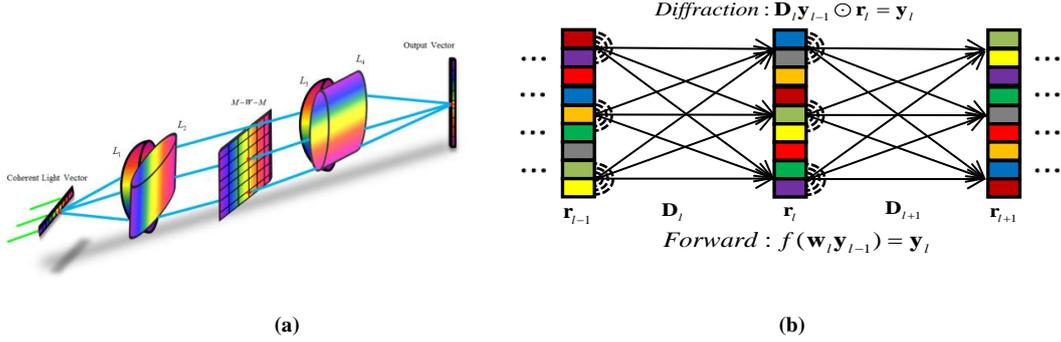

**(a)**  **(b)**

Fig.6 Schematic diagram of optical setup for diffractive neural network. **(a)** Optical intelligent unit for matrix-vector multiplication; **(b)** The graph of two-layer units for deep diffractive neural network

Deep diffractive neural network displaying in Fig 6(b) is build based on the analogous portrait of optical forward diffraction in free space. In deep complex neural network, a synapse collects the complex-value neurons, accomplishing by optical diffraction, and is further nonlinear activated with nonlinear optoelectronic response. Coherent modulation mechanism is introduced into nodes considering diffraction propagation matrix. The coherent diffraction and neural network in the $nth$ layers can be modulated to be an equivalence. $\mathbf{D}_n^\varepsilon$ is exploited to represent discrete coherent diffraction matrix.

*Coherent Diffraction Forward mode*

$$\mathbf{D}_n^\varepsilon(\mathbf{D}_{n-1}^\varepsilon(\mathbf{D}_{n-2}^\varepsilon\cdots(\mathbf{D}_2^\varepsilon(\mathbf{D}_1^\varepsilon\mathbf{y}_0\odot\mathbf{r}_1)\odot\mathbf{r}_2)\cdots\odot\mathbf{r}_{n-2})\odot\mathbf{r}_{n-1})\odot\mathbf{r}_n = \mathbf{y}_n$$

*Neural Network Forward model*

$$\mathbf{W}_n f_{n-1}(\mathbf{W}_{n-1} f_{n-2}(\mathbf{W}_{n-2}\cdots f_2(\mathbf{W}_2 f_1(\mathbf{W}_1\mathbf{y}_0)))) = \mathbf{y}_n$$

The modulation mechanism in the $\ell th$ layer is built as

$$\begin{cases} f_1(\mathbf{W}_1\mathbf{y}_0) = \mathbf{y}_1 \\ f_2(\mathbf{W}_2\mathbf{y}_1) = \mathbf{y}_2 \\ \vdots \\ f_n(\mathbf{W}_n\mathbf{y}_{n-1}) = \mathbf{y}_n \end{cases} \qquad \begin{cases} \mathbf{D}_1^\varepsilon\mathbf{y}_0\odot\mathbf{r}_1 = \mathbf{y}_1 \\ \mathbf{D}_2^\varepsilon\mathbf{y}_1\odot\mathbf{r}_2 = \mathbf{y}_2 \\ \vdots \\ \mathbf{D}_n^\varepsilon\mathbf{y}_{n-1}\odot\mathbf{r}_n = \mathbf{y}_n \end{cases}$$

$$f_\ell(\mathbf{W}_\ell\mathbf{y}_{\ell-1}) = \mathbf{D}_\ell^\varepsilon\mathbf{y}_{\ell-1}\odot\mathbf{r}_\ell \qquad (4)$$

$\odot$ denotes Hadamard product. $\varepsilon$ indicates the linear optical transform including Fourier、Fresnel、Fractional Fourier operator relating to diffractive distance. Diffraction distances are given as $z_n = d^2/\lambda N$ with the size in an area $d\times d$ with $N\times N$ pixels in each layer, $\lambda$ is the illuminating coherent wavelength. The coherent diffraction forward propagation is modulated to be the output optical signal in $nth$ layer through nonlinear activation , of which the input signal $\mathbf{y}_{n-1}$ is multiplied by the optical intelligence matrix learned with compatible learning, $\mathbf{r}_n$ is the forward modulation. Even optical architecture is built other than diffraction, such as scattering, if the photonic forward operator, like $\mathbf{D}_n^\varepsilon$, could be provided, modulation mechanism would be well done. $\mathbf{D}_n^\varepsilon$ in coherent diffraction propagation is deliberately demonstrated in supplementary materials 3.

We feed training samples in two-layer neural network in complex space, for preliminary applications, with random modulation [25-26] described in Figure 6(b), 1D signal is implemented here. The number of training samples are 50000, and the tested samples are 1000. In Figure 7(a)-(b), we show three types of training samples involving amplitude-only, phase-only, amplitude-phase, and the corresponding diffraction complex-amplitude optical field with angular spectrum [27]. The scale of the samples is 5mm with 512 sampling points under 632.8nm coherent illumination, and the scale of learned compatible matrix is set as 512×512 given at 0.78m diffraction distance between two neural layers. The mingled training samples consist of real-value and complex-value can also be successfully trained with compatible learning, which indicates the superiority of compatibility. The learned weights are related to the diffraction modulation mechanism, the random modulation can be simply obtained with $\mathbf{r}_n = \frac{f_n(\mathbf{W}_n \mathbf{y}_{n-1})}{\mathbf{D}_n^\varepsilon \mathbf{y}_{n-1}}$. The physical mechanism of forward diffraction is attainable for deep neural network in complex space. All the parameters are passive and fixed once the neural network is trained successfully, whereas the learned parameters are different derived from initialization. One of the tested results is presented in Fig. 7(d), the illuminated coherent wavelength and diffraction distance retains unchanged in the forward network test stage. During the iterations of training, the convergent curves reach nearly constant after 20 epochs in Figure 7(c), and presents good performance, the activation function tanh surpasses sigmoid in convergence speed under the same network environment. The diffraction distance and the coherent illumination wavelength could influence the random modulation for nodes or weight presenting a great potential on recognition occasions related to distance.

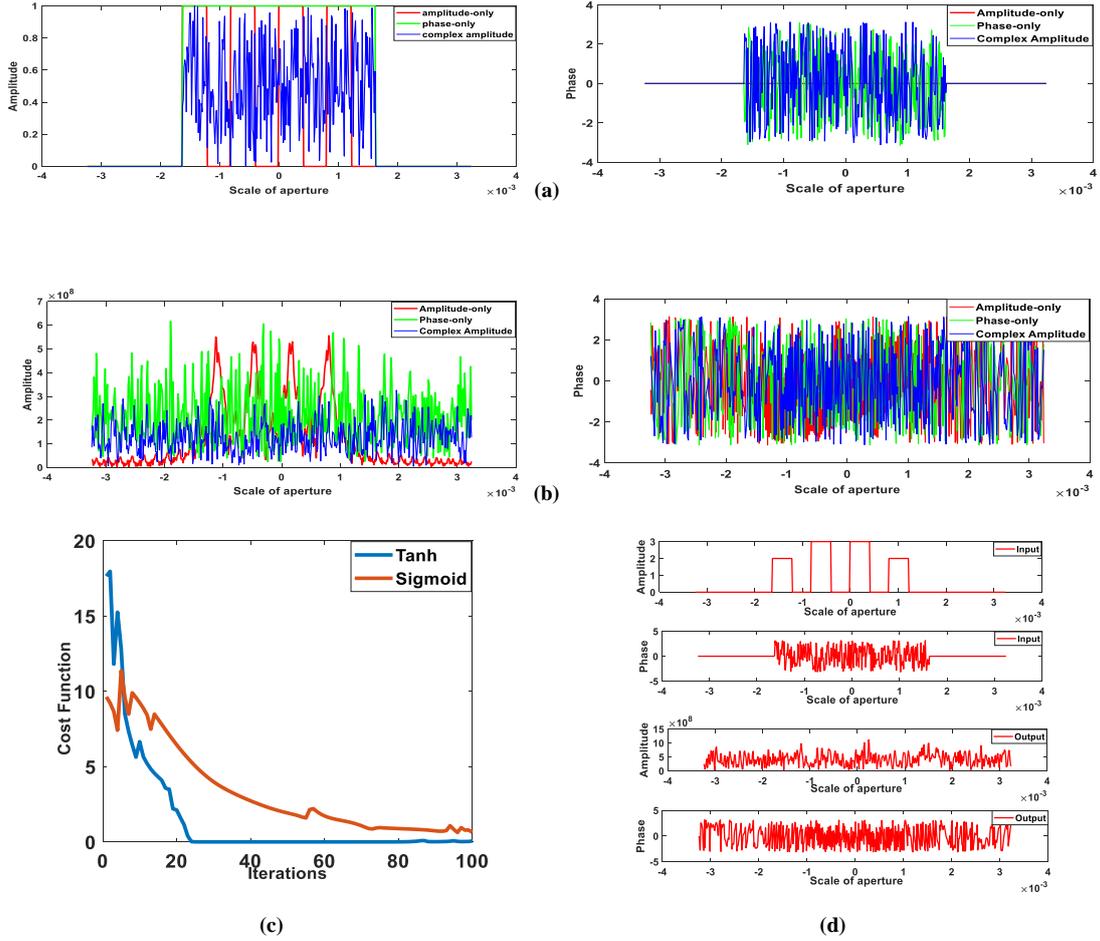

Fig.7 (**a**) Three types of the feeding diffractive samples for input layer, (**b**) The feeding diffractive samples for output layer, (**c**) The learning curves with different nonlinear activation, (**d**) The forward outputs for test samples.

*Intensity to intensity mapping with initialization in complex space*

For deep photonic neutral network, in general, the coherent optical field at the output layer involves amplitude and phase, optical interference is commonly introduced to record the phase. However, direct intensity recording without interference at the output layer is more convenient [28], phase lost would occur just in the output physical layer. The initialization in complex space with compatible learning can undertake the tasks. The intelligent matrix could be achievable statistically by feeding labels. Firstly, we present four types of random initialization in complex space within real-value XOR logical gate realized by 4×4 matrix in two-layer mapping for comparisons, phase-only, real and its mirror imaginary part, real and imaginary separation, in which the mean-values are set as zero with default variances in MATLAB. The learning curves are shown in Figure 8. Phase-only initialization presents step-descend convergent characteristic, real space initialization is faster than its mirrored imaginary part initialization on convergent speed. Particularly, the real and imaginary separate initialization outperforms the other types of initialization on convergent speed, and pure imaginary part initialization is also good. The overall output results within 200 iterations exhibit about 0.1% intensity errors compared to the labels. Observed from the convergence curves that the initialization in both real and complex space can accomplish the same prediction-oriented task in real space with compatible learning, proving that the passive intelligent matrix is completely dependent on the initialization under the same learning task in real space.

Secondly, the classification on the MNIST dataset is further trained, and sigmoid is chosen for judgment at output layer as well as tanh is chosen as the activation in hidden layer. The feeding labels are 2D intensity images without any phase information. We also present three types of uniformly random initialization in complex space to observe the performance of compatible learning within two-layer neural network. Some selected images from MNIST are displayed in Figure 9(a), and the number of entire samples for training is 60000, and learning rate is set as 2.0. The learning curves with different initialization are displayed in Figure 9(b). The holistic tendency of learning curves is roughly descended during iterations, the compatible learning is basically valid, even the performance is not very good due to the missing variance propagation portraits reinforced in initialization here [29-30]. The real and imaginary separate initialization primarily outperforms the performance of other types of initialization in complex space. In most cases, the convergence should be further enhanced by deeply considering the variance dynamic in initialization, promoting compatible learning available in big-data training.

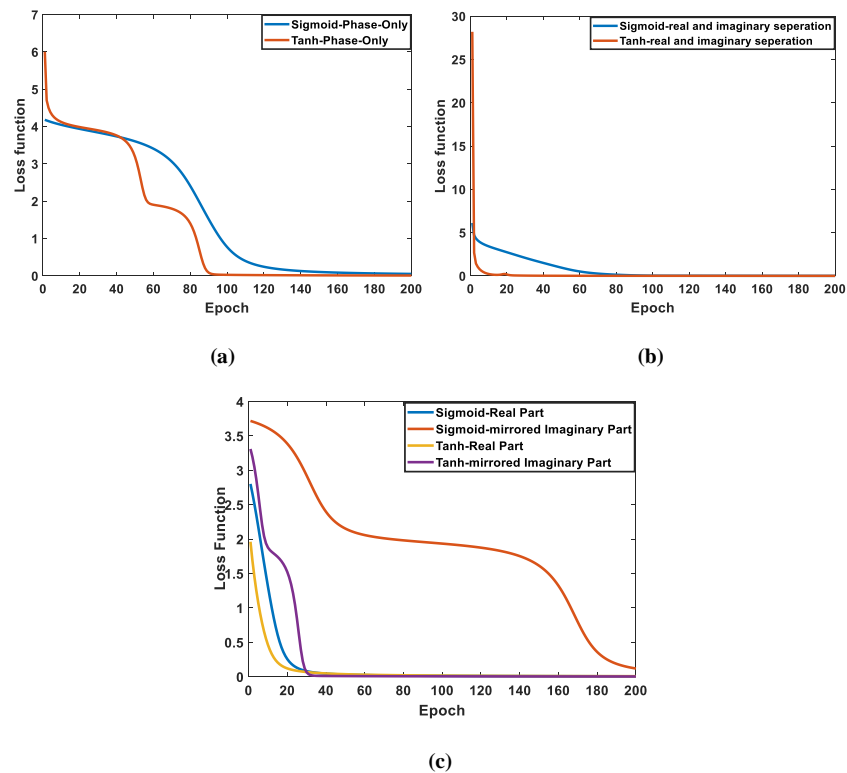

Fig.8 The initialization in complex space for intensity to intensity mapping with two-layer neural network implemented within real-value XOR logical gare through 4×4 weight. In each sub-figure of comparisons, the value of initialization for different activations are the same. **(a)** Phase-only initialization; **(b)** Real and imaginary separation initialization; **(c)** Real space and its mirrored imaginary part initialization.

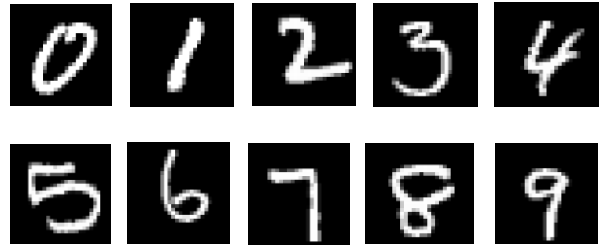

(a)

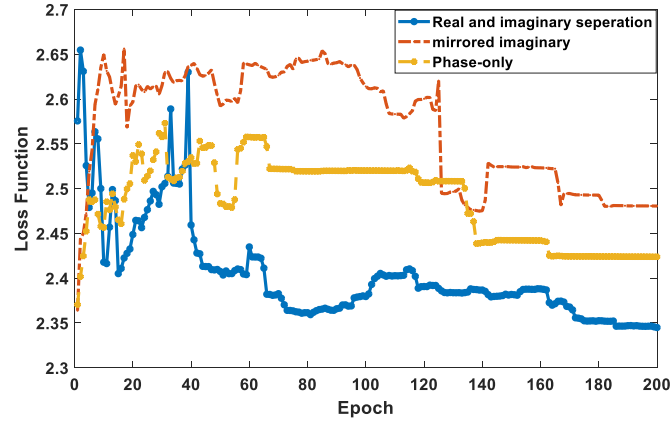

(b)

Fig.9  The initialization in complex space for intensity to intensity classified task with two-layer neural network implemented in MNIST dataset. **(a)** Selected samples exhibition; **(b)** Learning curves with compatible learning.

*Conclusions*

In the era of AI, the speed of feedforward realization for deep learning is of significance for real-time applications, and a series of different deep neural network architectures have been implemented on the oriented–electronic chip. However, photonic neural network has a great superiority of processing at speed of light, low power consumption as well as low latency, compatible learning in complex space provides a potential support for coherent optical field intelligent training, because of its competence on learning in a single channel, replacing the bumpy double channel training and activation that separated in real and imaginary part .

Compatible learning from BackPropogation directly adopts the complex-value entity to update the weight, the compatible condition ensures that some nonlinear activations in real space can be extended directly to complex space. Thus, deep neural network in real space and complex space can be effectively trained as a unity. The compatible condition, in which complex amplitude nonlinear activation can be realized by amplitude scaling and phase rotation, enhancing the practice of implementing photoelectric material as nonlinear activation, without modulation or manipulation in phase space. Moreover, weight update in complex space also presents indirect descriptions about the amplitude and phase corrections in each layer.

In conclusion, this article gives a description on compatible learning for deep photonic neural network. Two passive types of photonic neural network realized by optical diffraction and interference are trained by compatible learning. The compatible learning deserves promotion as an available protocol on deep neural network related to coherent wave physics. Additionally, ReLU in complex space could be interpreted as  an overlapping function squashed out considering compatible and C-R condition simultaneously.

**Funding**    This work was supported by National Natural Science Foundation of China (61805208), Program for Changjiang Scholars and Innovative Research Team in University (IRT-17R91), National Key Scientific Apparatus Development Project(2013YQ49087901).

# Supplementary materials

## 1

Relating to the separate double channel training in real and imaginary parts with real-value Backpropagation, a complex number multiplied by a complex number can be computed with multiplication of a 2×2 real-value submatrix and a 2×1 real-value vector as follow

$$W = A + iB, h = x + iy$$

$$Wh = \begin{bmatrix} A & -B \\ B & A \end{bmatrix} \begin{bmatrix} x \\ y \end{bmatrix} = \begin{bmatrix} \Re(Wh) \\ \Im(Wh) \end{bmatrix}$$

$$\begin{bmatrix} W_{11} & W_{12} \\ W_{21} & W_{22} \end{bmatrix} \begin{bmatrix} h_1 \\ h_2 \end{bmatrix} = \begin{bmatrix} A_{11} & -B_{11} & A_{12} & -B_{12} \\ B_{11} & A_{11} & B_{12} & A_{12} \\ A_{21} & -B_{21} & A_{22} & -B_{22} \\ B_{21} & A_{21} & B_{22} & -A_{22} \end{bmatrix} \begin{bmatrix} x_1 \\ y_1 \\ x_2 \\ y_2 \end{bmatrix} = \begin{bmatrix} (A_{11}x_1 - B_{11}y_1) + (A_{12}x_2 - B_{12}y_2) \\ (B_{11}x_1 + A_{11}y_1) + (B_{12}x_2 + A_{12}y_2) \\ (A_{21}x_1 - B_{21}y_1) + (A_{22}x_2 - B_{22}y_2) \\ (B_{21}x_1 + A_{21}y_1) + (B_{22}x_2 - A_{22}y_2) \end{bmatrix} = \begin{bmatrix} \Re(W_{11}h_1) + \Re(W_{12}h_2) \\ \Im(W_{11}h_1) + \Im(W_{12}h_2) \\ \Re(W_{21}h_1) + \Re(W_{22}h_2) \\ \Im(W_{21}h_1) + \Im(W_{22}h_2) \end{bmatrix}$$

The complex space neural network in a single layer feedforward format is a full connection. Training the real part with Real-Value Backpropagation in *kth* layer depending on the following feedforward format

$$f_k\left(\begin{bmatrix} A_{11}^k & -B_{11}^k & A_{12}^k & -B_{12}^k \\ A_{21}^k & -B_{21}^k & A_{22}^k & -B_{22}^k \end{bmatrix} \begin{bmatrix} x_1^k \\ y_1^k \\ x_2^k \\ y_2^k \end{bmatrix}\right) = f_k\left(\begin{bmatrix} C_1^k \\ C_2^k \end{bmatrix}\right)$$

After learning the above real parts of the weight, the learned weight is regularly stitched by a series of partition submatrix, which is expanded by diagonally crossing the learned real-value weight through the upsampling operation. However, as to complex neural network, the nonlinear activations in real and imaginary parts with the same real function don't conform to the compatible condition as well as C-R condition, which loses a certain degree of availability for effective training in deep complex neural network. The activation in real-value Backpropagation for complex-value weight should be enforced to be compatible with the requirement for nonlinear activation in complex space. So does the random initialization. It is passive and negative. Additionally, photonic neural network realization in such a format become tough because coherent optical field in the form of complex-amplitude is forward propagated and actived nonlinearly as an entity.

## 2

The Mach-Zehnder Interferometer (MZI) is a photonic device with high phase-shift sensitivity [1]. Each MZI unit consists of four sections, in which there are two 3dB beam-splitter and two phase-shifter arms. The function of 3dB beam-splitter and phase-shifter are described using transfer matrix represented by matrices M and $T_\theta$, $T_\phi$, respectively. Fig.1 is a schematic diagram of the structure of a designed 4×4 meshing interferometers [2].

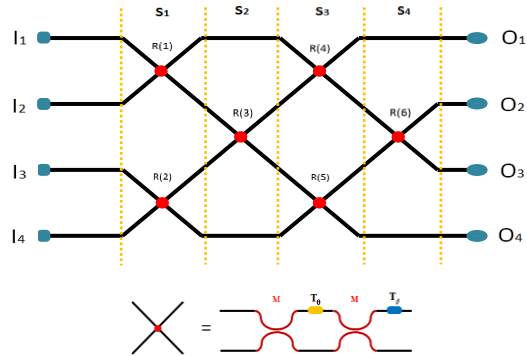

**Figure 1**. *Schematic diagram of the structure of a designed 4×4 meshing interferometers, Single-stage MZI structure comprising two 3dB beam-splitter and two phase-shift.*

Wherein, the matrix M of the couplers is expressed

$$M = \frac{1}{\sqrt{2}} \begin{bmatrix} 1 & i \\ i & 1 \end{bmatrix}$$

Two tunable phase-shifter are the key parts of the MZI unit. The purpose of the optical signal modulation can be achieved by tuning the two phase-shifter. The matrix $T_\theta$, $T_\phi$ of the two phase-shifter are given as

$$T_\phi = \begin{bmatrix} e^{i\phi} & 0 \\ 0 & 1 \end{bmatrix}, \quad T_\theta = \begin{bmatrix} e^{i\theta} & 0 \\ 0 & 1 \end{bmatrix}$$

As an MZI unit, its transfer matrix can be written as

$$R = T_\phi M T_\theta M = \frac{1}{2} \begin{bmatrix} e^{i\phi} & 0 \\ 0 & 1 \end{bmatrix} \begin{bmatrix} 1 & i \\ i & 1 \end{bmatrix} \begin{bmatrix} e^{i\theta} & 0 \\ 0 & 1 \end{bmatrix} \begin{bmatrix} 1 & i \\ i & 1 \end{bmatrix}$$

$$= ie^{i\theta/2} \begin{bmatrix} e^{i\phi} \sin(\theta/2) & e^{i\phi} \cos(\theta/2) \\ \cos(\theta/2) & -\sin(\theta/2) \end{bmatrix}$$

In order to realize optically the N×N weight between the layer by layer, the matrix dimensions can be meshed designedly by integrating MZI units. For the nth MZI unit, the transmission matrix is:

$$R(n) = T_{\phi_n} M T_{\theta_n} M = ie^{i\theta_n/2} \begin{bmatrix} e^{i\phi_n} \sin(\theta_n/2) & e^{i\phi_n} \cos(\theta_n/2) \\ \cos(\theta_n/2) & -\sin(\theta_n/2) \end{bmatrix}$$

The relationship for the 4×4 optical intelligent weight can be written as $O = S_4 S_3 S_2 S_1 I$

We define S as $\quad S = S_4 S_3 S_2 S_1$

Where

$$S = \begin{bmatrix} 1 & & \\ & R_6 & \\ & & 1 \end{bmatrix} \begin{bmatrix} R_4 & \\ & R_5 \end{bmatrix} \begin{bmatrix} 1 & & \\ & R_3 & \\ & & 1 \end{bmatrix} \begin{bmatrix} R_1 & \\ & R_2 \end{bmatrix}$$

S implements the intelligent weight by manipulating phase-shifter parameters, and there are non-zero elements appearing in S. It is not a unitary matrix and can be unitary if regularization is enforced on backpropagation in complex space.

Once the passive weight $W$ is learned with different initialization, we can determine phase-shift value by element equivalence within $S = W$.

$$W_{1,1} = e^{i[\phi_1 + \phi_4 + (\theta_1 + \theta_4)/2 + \pi]} \sin\frac{\theta_1}{2} \sin\frac{\theta_4}{2} + e^{i[\phi_3 + \phi_4 + (\theta_1 + \theta_3 + \theta_4)/2 + 3\pi/2]} \cos\frac{\theta_1}{2} \cos\frac{\theta_4}{2} \sin\frac{\theta_3}{2}$$

$$W_{2,1} = e^{i[\phi_1 + \phi_6 + (\theta_1 + \theta_4 + \theta_6)/2 + 3\pi/2]} \sin\frac{\theta_1}{2} \sin\frac{\theta_6}{2} \cos\frac{\theta_4}{2} - e^{i[\phi_3 + \phi_6 + (\theta_1 + \theta_3 + \theta_4 + \theta_6)/2 + 2\pi]} \sin\frac{\theta_3}{2} \sin\frac{\theta_4}{2} \sin\frac{\theta_6}{2} \cos\frac{\theta_1}{2} + e^{i[\phi_5 + \phi_6 + (\theta_1 + \theta_3 + \theta_5 + \theta_6)/2 + 2\pi]} \sin\frac{\theta_5}{2} \cos\frac{\theta_1}{2} \cos\frac{\theta_3}{2} \cos\frac{\theta_6}{2}$$

$$W_{3,1} = e^{i[\phi_1 + (\theta_1 + \theta_4 + \theta_6)/2 + 3\pi/2]} \sin\frac{\theta_1}{2} \cos\frac{\theta_6}{2} \cos\frac{\theta_4}{2} - e^{i[\phi_3 + (\theta_1 + \theta_3 + \theta_4 + \theta_6)/2 + 2\pi]} \sin\frac{\theta_3}{2} \sin\frac{\theta_4}{2} \cos\frac{\theta_1}{2} \cos\frac{\theta_6}{2} - e^{i[\phi_5 + (\theta_1 + \theta_3 + \theta_5 + \theta_6)/2 + 2\pi]} \sin\frac{\theta_1}{2} \sin\frac{\theta_5}{2} \sin\frac{\theta_6}{2} \cos\frac{\theta_3}{2}$$

$$W_{4,1} = e^{i[(\theta_1 + \theta_3 + \theta_5)/2 + 3\pi/2]} \cos\frac{\theta_1}{2} \cos\frac{\theta_3}{2} \cos\frac{\theta_5}{2}$$

$$W_{1,2} = e^{i[\phi_1 + \phi_4 + (\theta_1 + \theta_4)/2 + \pi]} \sin\frac{\theta_4}{2} \cos\frac{\theta_1}{2} - e^{i[\phi_3 + \phi_4 + (\theta_1 + \theta_3 + \theta_4)/2 + 3\pi/2]} \sin\frac{\theta_1}{2} \sin\frac{\theta_3}{2} \cos\frac{\theta_4}{2}$$

$$W_{2,2} = e^{i[\phi_1 + \phi_6 + (\theta_1 + \theta_4 + \theta_6)/2 + 3\pi/2]} \sin\frac{\theta_6}{2} \cos\frac{\theta_1}{2} \cos\frac{\theta_4}{2} + e^{i[\phi_3 + \phi_6 + (\theta_1 + \theta_3 + \theta_4 + \theta_6)/2 + 2\pi]} \sin\frac{\theta_1}{2} \sin\frac{\theta_3}{2} \sin\frac{\theta_4}{2} \sin\frac{\theta_6}{2} - e^{i[\phi_5 + \phi_6 + (\theta_1 + \theta_3 + \theta_5 + \theta_6)/2 + 2\pi]} \sin\frac{\theta_1}{2} \sin\frac{\theta_5}{2} \cos\frac{\theta_3}{2} \cos\frac{\theta_6}{2}$$

$$W_{3,2} = e^{i[\phi_1 + (\theta_1 + \theta_4 + \theta_6)/2 + 3\pi/2]} \cos\frac{\theta_1}{2} \cos\frac{\theta_4}{2} \cos\frac{\theta_6}{2} + e^{i[\phi_3 + (\theta_1 + \theta_3 + \theta_4 + \theta_6)/2 + 2\pi]} \sin\frac{\theta_1}{2} \sin\frac{\theta_3}{2} \sin\frac{\theta_4}{2} \cos\frac{\theta_6}{2} + e^{i[\phi_5 + (\theta_1 + \theta_3 + \theta_5 + \theta_6)/2 + 2\pi]} \sin\frac{\theta_1}{2} \sin\frac{\theta_5}{2} \sin\frac{\theta_6}{2} \cos\frac{\theta_3}{2}$$

$$W_{4,2} = -e^{i[(\theta_1 + \theta_3 + \theta_5)/2 + 3\pi/2]} \sin\frac{\theta_1}{2} \cos\frac{\theta_3}{2} \cos\frac{\theta_5}{2}$$

$$W_{1,3} = e^{i(\phi_2 + \phi_4 + (\theta_2 + \theta_3 + \theta_4)/2 + 3\pi/2)} \sin\frac{\theta_2}{2} \cos\frac{\theta_3}{2} \cos\frac{\theta_4}{2}$$

$$W_{2,3} = -e^{i(\phi_2 + \phi_3 + \phi_6 + (\theta_2 + \theta_3 + \theta_4 + \theta_6)/2 + 2\pi)} \sin\frac{\theta_2}{2} \sin\frac{\theta_4}{2} \sin\frac{\theta_6}{2} \cos\frac{\theta_3}{2} - e^{i(\phi_2 + \phi_5 + \phi_6 + (\theta_2 + \theta_3 + \theta_5 + \theta_6)/2 + 2\pi)} \sin\frac{\theta_2}{2} \sin\frac{\theta_3}{2} \sin\frac{\theta_5}{2} \cos\frac{\theta_6}{2} + e^{i(\phi_6 + \phi_6 + (\theta_2 + \theta_5 + \theta_6)/2 + 3\pi/2)} \cos\frac{\theta_2}{2} \cos\frac{\theta_5}{2} \cos\frac{\theta_6}{2}$$

$$W_{3,3} = -e^{i(\phi_2 + \phi_3 + (\theta_2 + \theta_3 + \theta_4 + \theta_6)/2 + 2\pi)} \sin\frac{\theta_2}{2} \sin\frac{\theta_4}{2} \cos\frac{\theta_3}{2} \cos\frac{\theta_6}{2} + e^{i(\phi_5 + (\theta_2 + \theta_3 + \theta_5 + \theta_6)/2 + 2\pi)} \sin\frac{\theta_2}{2} \sin\frac{\theta_3}{2} \sin\frac{\theta_5}{2} \sin\frac{\theta_6}{2} - e^{i(\phi_6 + (\theta_2 + \theta_5 + \theta_6)/2 + 3\pi/2)} \sin\frac{\theta_6}{2} \cos\frac{\theta_2}{2} \cos\frac{\theta_5}{2}$$

$$W_{4,3} = -e^{i(\phi_2 + (\theta_2 + \theta_3 + \theta_5)/2 + 3\pi/2)} \sin\frac{\theta_2}{2} \sin\frac{\theta_3}{2} \cos\frac{\theta_5}{2} - e^{i((\theta_2 + \theta_5)/2 + \pi)} \sin\frac{\theta_5}{2} \cos\frac{\theta_2}{2}$$

$$W_{1,4} = e^{i(\phi_2 + \phi_3 + \phi_4 + (\theta_2 + \theta_3 + \theta_4)/2 + 3\pi/2)} \cos\frac{\theta_2}{2} \cos\frac{\theta_3}{2} \cos\frac{\theta_4}{2}$$

$$W_{2,4} = -e^{i(\phi_2+\phi_5+\phi_6+(\theta_2+\theta_3+\theta_4+\theta_6)/2+2\pi)} \sin\frac{\theta_4}{2}\sin\frac{\theta_6}{2}\cos\frac{\theta_2}{2}\cos\frac{\theta_3}{2} - e^{i(\phi_5+\phi_6+(\theta_2+\theta_3+\theta_5+\theta_6)/2+2\pi)} \sin\frac{\theta_3}{2}\sin\frac{\theta_5}{2}\cos\frac{\theta_2}{2}\cos\frac{\theta_6}{2} - e^{i(\phi_6+(\theta_2+\theta_5+\theta_6)/2+3\pi/2)} \sin\frac{\theta_2}{2}\cos\frac{\theta_5}{2}\cos\frac{\theta_6}{2}$$

$$W_{3,4} = -e^{i(\phi_2+\phi_5+(\theta_2+\theta_3+\theta_4+\theta_6)/2+2\pi)} \sin\frac{\theta_4}{2}\cos\frac{\theta_2}{2}\cos\frac{\theta_3}{2}\cos\frac{\theta_6}{2} + e^{i(\phi_2+\phi_5+(\theta_2+\theta_3+\theta_5+\theta_6)/2+2\pi)} \sin\frac{\theta_3}{2}\sin\frac{\theta_5}{2}\sin\frac{\theta_6}{2}\cos\frac{\theta_2}{2} + e^{i(\phi_5+(\theta_2+\theta_5+\theta_6)/2+3\pi/2)} \sin\frac{\theta_2}{2}\sin\frac{\theta_6}{2}\cos\frac{\theta_5}{2}$$

$$W_{4,4} = -e^{i(\phi_2+(\theta_2+\theta_3+\theta_5)/2+3\pi/2)} \sin\frac{\theta_3}{2}\cos\frac{\theta_2}{2}\cos\frac{\theta_5}{2} + e^{i((\theta_2+\theta_5)/2+\pi)} \sin\frac{\theta_2}{2}\sin\frac{\theta_5}{2}$$

$\theta_i$, $\phi_i$ are the phase-shift parameters in $ith$ MZ interferometer.

### 3

The matrix representation for coherent diffraction propagation [3] is given as

$$D_\varepsilon = \frac{1}{\sqrt{N}} \begin{bmatrix} 1 & W^{(1/2)m^2} & \cdots & W^{(1/2)(N-1)^2} \\ W^{(1/2)n^2(1-\varepsilon^2)} & W^{(1/2)[(n-m)^2-n^2\varepsilon^2]} & \cdots & W^{(1/2)\{[n-(N-1)]^2-n^2\varepsilon^2\}} \\ \vdots & \vdots & \vdots & \vdots \\ W^{(1/2)(N-1)^2(1-\varepsilon^2)} & W^{(1/2)\{[(N-1)-m]^2-(N-1)^2\varepsilon^2\}} & \cdots & W^{-(1/2)(N-1)^2\varepsilon^2} \end{bmatrix}$$

$W = \exp(j\frac{2\pi}{N})$, $\varepsilon = \sin\phi$ is the fractional factor, $\varepsilon = 0$ means discrete Fresnel matrix, $\varepsilon = 1$ means discrete Fourier matrix.